\documentclass[12pt]{article}
\usepackage{amsmath}
\usepackage{times}
\usepackage{bm}
\usepackage{natbib}
\usepackage[plain,noend]{algorithm2e}

\usepackage{float}
\usepackage{graphicx}
\usepackage{comment}
\graphicspath{ {figs/} }
\usepackage[english]{babel}
\usepackage{dsfont}
\usepackage[OT1]{fontenc}
\usepackage[utf8]{inputenc}
\usepackage[inline,shortlabels]{enumitem}
\usepackage{lmodern}
\usepackage{color}
\usepackage{booktabs,caption}

\usepackage{refcount,nameref,zref-xr,zref-user} 
\zxrsetup{toltxlabel} 
\usepackage{url}

\usepackage{mathtools}
\usepackage{amsfonts}
\usepackage{bbm}

\makeatletter
\@ifclassloaded{biometrika}{%
  \makeatletter
  \renewcommand{\algocf@captiontext}[2]{#1\algocf@typo. \AlCapFnt{}#2}
  \def\@algocf@capt@plain{top}
  \renewcommand{\algocf@makecaption}[2]{%
    \addtolength{\hsize}{\algomargin}%
    \sbox\@tempboxa{\algocf@captiontext{#1}{#2}}%
    \ifdim\wd\@tempboxa >\hsize
      \hskip .5\algomargin%
      \parbox[t]{\hsize}{\algocf@captiontext{#1}{#2}}
    \else%
      \global\@minipagefalse%
      \hbox to\hsize{\box\@tempboxa}
    \fi%
    \addtolength{\hsize}{-\algomargin}%
  }
  \makeatother

  \addtolength\topmargin{35pt}

  \newcommand{\titlepaper}{Nonparametric inverse probability weighted
    estimators based on the highly adaptive lasso}

  \jname{Biometrika}

  \markboth{A.~Ertefaie, N.S.~Hejazi, M.J.~van der Laan}{Inverse probability
    weighted estimators based on the highly adaptive lasso}

  \newcommand{\authorlist}{
      \author{A.~ERTEFAIE}
      \affil{Department of Biostatistics and Computational Biology, University
        of Rochester\\ 265 Crittenden Boulevard, Rochester, New York 14642-0630
      \email{ashkan\_ertefaie@urmc.rochester.edu}}

      \author{N.S.~HEJAZI}
      \affil{Graduate Group in Biostatistics, University of California,
      Berkeley\\ 2121 Berkeley Way, Berkeley, CA 94720-7360
      \email{nhejazi@berkeley.edu}}

      \author{M.J.~VAN DER LAAN}
      \affil{Division of Epidemiology \& Biostatistics, University of
      California, Berkeley\\ 2121 Berkeley Way, Berkeley, CA 94720-7360
      \email{laan@berkeley.edu}}
  }
}{%
  \usepackage{arxiv}
  \usepackage{setspace}
  \usepackage{multirow}
  \usepackage[colorlinks,citecolor=blue,urlcolor=blue]{hyperref}

  \usepackage{amsthm}
  \usepackage{amsxtra}
  \usepackage{amssymb}
  \usepackage{commath}

  {
    \newtheorem{theorem}{Theorem}
    \newtheorem{lemma}{Lemma}
    \newtheorem{assumption}{Assumption}
  }

  \newcommand{\titlepaper}{Nonparametric inverse probability weighted
    estimators based on the highly adaptive lasso}

  \newcommand{\authorlist}{
    Ashkan Ertefaie \\
    Department of Biostatistics and Computational Biology, \\
    University of Rochester \\
    \texttt{ashkan\_ertefaie@urmc.rochester.edu} \\
    \And
    Nima S.~Hejazi \\
    Division of Epidemiology \& Biostatistics, \\
    School of Public Health, and \\
    Center for Computational Biology, \\
    University of California, Berkeley \\
    \texttt{nhejazi@berkeley.edu} \\
   \And
    Mark J.~{van der Laan} \\
    Division of Epidemiology \& Biostatistics, \\
    School of Public Health, and \\
    Department of Statistics, \\
     University of California, Berkeley \\
    \texttt{laan@berkeley.edu} \\
  }
}
\makeatother

\AtEndDocument{\refstepcounter{theorem}\label{finalthm}}
\AtEndDocument{\refstepcounter{lemma}\label{finallemma}}
\AtEndDocument{\refstepcounter{equation}\label{finaleq}}

\def\E{\mathbb{E}}
\def\M{\mathcal{M}}
\DeclareMathOperator{\expit}{expit}

\DeclareMathOperator{\logit}{logit}

\def\R{\mathbbmss{R}}
\def\D{\mathbbmss{D}}

\def\I{\mathbbmss{1}}

\def\I{\mathbbmss{1}}

\DeclareMathOperator*{\argmin}{argmin}

\DeclarePairedDelimiterX{\inp}[2]{\langle}{\rangle}{#1,#2}

\zexternaldocument*[supp:]{sm}
\title{\titlepaper}

\makeatletter
\@ifclassloaded{biometrika}{%
  \authorlist
}{%
  \author{\authorlist}
  \date{\today}
}
\makeatother

\begin{document}
\maketitle

\begin{abstract}
Inverse probability weighted estimators are the oldest and potentially most
commonly used class of procedures for the estimation of causal effects. By
adjusting for selection biases via a weighting mechanism, these procedures
estimate an effect of interest by constructing a pseudo-population in which
selection biases are eliminated. Despite their ease of use, these estimators
require the correct specification of a model for the weighting mechanism, are
known to be inefficient, and suffer from the curse of dimensionality. We propose
a class of nonparametric inverse probability weighted estimators in which the
weighting mechanism is estimated via undersmoothing of the highly adaptive
lasso, a nonparametric regression function proven to converge at $n^{-1/3}$-rate
to the true weighting mechanism. We demonstrate that our estimators are
asymptotically linear with variance converging to the nonparametric efficiency
bound. Unlike doubly robust estimators, our procedures require neither
derivation of the efficient influence function nor specification of the
conditional outcome model. Our theoretical developments have broad implications
for the construction of efficient inverse probability weighted estimators in
large statistical models and a variety of problem settings. We assess the
practical performance of our estimators in simulation studies and demonstrate
use of our proposed methodology with data from a large-scale epidemiologic
study.
\end{abstract}

\section{Introduction}\label{intro}

Inverse probability weighted estimators have been widely used in a diversity of
fields, as inverse probability weighting allows for the adjustment of selection
biases by the assignment of weights (i.e., based on propensity scores) to
observational units such that a pseudo-population mimicking the target
population is generated. The construction of inverse probability weighted
estimators is relatively straightforward, as the only nuisance parameter that
must be estimated is the propensity score. Owing in part to the ease with which
inverse probability weighted estimators may be constructed, their application
has been frequent in causal inference~\citep[e.g.,][]{robins2000marginal},
missing data~\citep[e.g.,][]{robins1994estimation}, and survival
analysis~\citep[e.g.,][]{hernan2000marginal,tsiatis2007semiparametric}.

While inverse probability weighting may easily be implemented and is appropriate
for use in a variety of problem settings, the resultant estimators face several
disadvantages. Unfortunately, such estimators require a correctly specified
estimate of the propensity score to produce consistent estimates of the target
parameter and can be inefficient in certain settings (e.g., randomized
controlled trials). What is more, inverse probability weighted estimators suffer
from the curse of dimensionality, as their rate of convergence depends entirely
on the convergence rate of the postulated model for the propensity score. This
latter requirement has proven a significant obstacle to investigators wishing to
use data adaptive techniques in the estimation of propensity scores. To overcome
these significant shortcomings,~\citet{vdl2014targeted} proposed the
\textit{targeted} inverse probability weighted estimator, which facilities the
use of data adaptive techniques in estimating the relevant weight functions.
While the targeted estimator is asymptotically linear, it has been shown to
suffer from issues of irregularity~\citep{vdl2014targeted}. Alternatively,
doubly robust estimation procedures, which are based on constructing models for
both the propensity score and the outcome mechanism~\citep{bang2005doubly}, were
proposed. Doubly robust estimators are consistent for the target parameter when
either one of the two nuisance parameters is consistently estimated; moreover,
such estimators are efficient when both nuisance parameter estimators are
correctly specified~\citep{rotnitzky1998semiparametric, vdl2003unified}. While
doubly robust estimators allow two opportunities for consistent estimation,
their performance depends critically on the choice of estimators of these
nuisance parameters. When finite-dimensional models are used to estimate the two
nuisance parameters, doubly robust estimators may perform poorly, due to the
possibility of model misspecification in either of the nuisance parameter
estimators~\citep{kang2007demystifying, cao2009improving, vermeulen2015bias,
vermeulen2016data}. Although doubly robust procedures facilitate the use of data
adaptive techniques for modeling nuisance parameters, the resultant estimator
can be irregular with large bias and a slow rate of convergence when either of
the nuisance parameters is inconsistently estimated. To ease such issues,
\citet{vdl2014targeted} proposed a targeted doubly robust estimator that does
not suffer from the irregularity issue; the properties of this estimation
procedure were subsequently further investigated by~\citet{benkeser2017doubly}.

While many data adaptive regression techniques have been shown to provide
consistent estimates in flexible models, establishing the rate of convergence
for such approaches is often a significant obstacle. Among such approaches, the
highly adaptive lasso stands out for its ability to flexibly estimate arbitrary
functional forms with a fast rate of convergence under relatively mild
conditions. The highly adaptive lasso is a nonparametric regression function
that minimizes a loss-specific empirical risk over linear combinations of
indicator basis functions under the constraint that the sum of the absolute
value of the coefficients is bounded by a constant~\citep{vdl2017generally,
vdl2017uniform}. Letting the space of the functional parameter be a subset of
the set of c\`{a}dl\`{a}g (right-hand continuous with left-hand limits)
functions with sectional variation norm bounded by a finite constant,
\citet{vdl2017generally} showed that the highly adaptive lasso estimator
converges to the true value at a rate faster than $n^{-1/4}$, regardless of
dimensionality $d$. \citet{bibaut2019fast} subsequently improved this
convergence rate to $n^{-1/3} \log(n)^{d/2}$. Unlike most existing data adaptive
techniques that require local smoothness assumptions on the true functional
form, the finite sectional variation norm assumption imposed by the highly
adaptive lasso constitutes only a (less restrictive) global smoothness
assumption, making it a powerful approach for use in a variety of settings.

We show that inverse probability weighted estimators can be asymptotically
(nonparametric) efficient when the propensity score is estimated using a highly
adaptive lasso estimator tuned in a particular manner. Specifically, we show
that undersmoothing of the highly adaptive lasso allows for the resultant
inverse probability weighted estimator of the target parameter to be
asymptotically linear and a solution to an appropriate efficient influence
function equation. In the typical construction of highly adaptive lasso
estimators, cross-validation is used to determine the sectional variation norm
of the underlying functional parameter. By contrast, undersmoothing of the
highly adaptive lasso allows for a sectional variation norm greater than the
choice made by the global cross-validation selector to be used. A significant
challenge arises in finding a suitable choice of sectional variation norm ---
one that results in sufficient undersmoothing while simultaneously avoiding
overfitting. We provide theoretical conditions under which the desired degree of
undersmoothing may be achieved; moreover, we supplement our theoretical
investigations by providing practical guidance on how appropriate choices may be
made for the required tuning parameters in practice. Our proposed approach
obviates many of the challenges associated with the current methods of choice,
namely
\begin{enumerate}[label=(\roman*),leftmargin=1cm]
  \item in contrast with standard inverse probability weighted estimators, our
      estimators do not suffer from an \textit{asymptotic} curse of
      dimensionality, allowing the construction of asymptotically efficient
      estimators;
  \item in contrast with targeted inverse probability weighted estimators, our
      estimators do not suffer from potential issues of irregularity; and
  \item in contrast with typical doubly robust and efficient estimators, our
      estimators rely on only a single nuisance parameter and may be formulated
      without derivation of the efficient influence function.
\end{enumerate}

\section{Preliminaries}\label{prelim}

\subsection{Problem formulation, notation, and target parameter}\label{setup}

Consider data generated by typical cohort sampling: let $O = (W,A,Y) \sim P_0
\in \M$ be the data on a given observational unit, where $P_0$, the distribution
of $O$, lies in the nonparametric model $\M$. $W \in \mathcal{W}$ constitutes
baseline covariates measured prior to treatment $A \in \{0,1\}$; $Y$ is an
outcome of interest. Suppose we observe a sample of $n$ independent and
identically distributed units $O_1, \ldots O_n$, whose empirical distribution we
denote $P_n$. We let $Pf = \int f(o) dP(o)$ for a given function $f(o)$ and
distribution $P$, denoting by $\E_P$ expectations with respect to $P$. Let $G
: \M \rightarrow \mathcal{G}$ be a functional nuisance parameter where
$\mathcal{G}=\{G(P): P \in \M\}$. We use $G \equiv G(P) \equiv G(P)(A \mid W)$
to denote the treatment mechanism under an arbitrary distribution $P \in \M$. We
refer to the treatment mechanism under the true data-generating distribution as
$G_0$, that is, $G_0 = G(P_0)$. Letting $Y^a$ be the potential outcome that
would have been observed under the intervention
$\text{do}(A=a)$~\citep{pearl2000causality}, we define the full data unit $X$ as
$X = \{W, Y^0, Y^1\} \sim P_X \in \M^F$. A common parameter of interest is the
mean counterfactual outcome under treatment, i.e., $\Psi^F(P_X)
= \E_{P_X}(Y^1)$, where $\Psi^F: \M^F \rightarrow \R$ and $\M^F$ is the
nonparametric model for the full data $X$. While we present our results in the
context of this target parameter, we stress that our developments extend to any
arbitrary $a \in \mathcal{A}$ without loss of generality. Define the
corresponding full data canonical gradient $D^F(X, \Psi^F) = \{Y^1
- \Psi^F(P_X)\}$ and allow $\prod$ to be a projection operator in the Hilbert
space $L_0^2(P)$ with inner product $\inp{h_1}{h_2} = \E_P(h_1h_2)$. To identify
the causal effect of interest, we assume consistency (i.e., $Y_i=Y^{A_i}$) and
no unmeasured confounding or strong ignorability (i.e., $A \perp Y^a \mid
W$)~\citep{pearl2000causality, imbens2015causal, hernan2020causal}. Consistency
links the potential outcomes to those observed while strong ignorability is
a particular case of the randomization (i.e., coarsening at random) assumption.

\subsection{Inverse probability weighted mapping}\label{ipw_mapping}

As we only observe one of the potential outcomes for each unit, we define
an inverse probability weighted mapping of $D^F(X,\Psi^F)$ so as to estimate the
target parameter $\Psi(P_X)$ using the observed data:
\begin{equation*}
  U_G(O; \Psi) = \frac{A Y}{G(1 \mid W)} - \Psi(P),
\end{equation*}
where $\Psi: \M \rightarrow \R$. Here, $\Psi(P) = \E_{P}(Y^1) = \E_P \{\E_P(Y
\mid A = 1, W)\}$. Under the standard identification assumptions noted in
section~\ref{setup}, $\E_P \{U_G(O; \Psi) \mid X \}= D^F(X,\Psi^F)$. Under
coarsening at random, the tangent space of $G$ may be defined as
$T_{\text{CAR}} = \{\eta(A,{W}): \E_{P_0} \{\eta(A,{W}) \mid W\} = 0\}$. The
canonical gradient of $\Psi$ at a distribution $P \in \M$ is
\begin{equation*}
  D^{\star}(P) = U_G(O; \Psi) - D_{\text{CAR}}(P),
\end{equation*}
where $D_{\text{CAR}}(P) = \prod \{U_G(\Psi) \mid
T_{\text{CAR}}\}$~\citep{robins1994estimation, vdl2003unified}.
Following~\citet{vdl2003unified}, we have that $\prod \{U_G( \Psi) \mid
T_{\text{CAR}}\} = \E_P\{U_G(O; \Psi) \mid A=1, {W}\} - \E_P\{U_G(O; \Psi)
\mid {W}\}$, which may equivalently be expressed
\begin{equation*}
  D_{\text{CAR}}(P) = \frac{A - G(A \mid W)}{G(A \mid W)} Q(1,W),
\end{equation*}
where $Q(1,{W}) = \E_P(Y \mid A = 1, W)$ is the conditional mean outcome.

\subsection{The highly adaptive lasso estimator}\label{hal}

The highly adaptive lasso is a nonparametric regression function with the
capability to estimate infinite-dimensional functional parameters at
a near-parametric rate under relatively mild
assumptions~\citep{vdl2017generally, vdl2017uniform}. \citet{benkeser2016highly}
first demonstrated the utility of this estimator in extensive simulation
experiments. In its zeroth-order formulation, the highly adaptive lasso
estimator constructs a linear combination of indicator basis functions to
minimize the expected value of a loss function under the constraint that the
$L_1$-norm of the vector of coefficients is bounded by a finite constant
matching the sectional variation norm.

Let $\D[0,\tau]$ be the Banach space of $d$-variate real-valued c\`{a}dl\`{a}g
functions on a cube $[0,\tau] \in \R^d$. For each function $f \in \D[0,\tau]$,
define the supremum norm as $\lVert f \rVert_{\infty}=\sup_{w \in [0,\tau]}
\lvert f(w) \rvert$. For any subset $s$ of $\{0,1,\ldots,d\}$, partition
$[0,\tau]$ in $\{0\} \{ \cup_s (0_s,\tau_s]\}$ and define the sectional
variation norm of a given $f$ as
\begin{equation*}
  \lVert f \rVert^{\star}_\nu = \lvert f(0) \rvert +\sum_{s
  \subset\{1,\ldots,d\}} \int_{0_s}^{\tau_s} \lvert df_s(u_s) \rvert,
\end{equation*}
where the sum is over all subsets of the coordinates $\{0,1,\ldots,d\}$. For a
given subset $s \subset \{0,1,\ldots,d\}$, define $u_s = (u_j : j \in s)$ and
$u_{-s}$ as the complement of $u_s$. Then, $f_s: [0_s, \tau_s] \rightarrow \R$,
defined as $f_s(u_s) = f(u_s,0_{-s})$. Thus, $f_s(u_s)$ is a section of $f$ that
sets the components in the complement of $s$ to zero, i.e, varying only along
components in $u_s$.

Under the assumption that our nuisance functional parameter $G \in \D[0,\tau]$
has finite sectional variation norm, $\logit G$ may be
represented~\citep{gill1995inefficient}:
\begin{align}\label{eq:hal1}
  \logit G(w) &= \logit G(0) + \sum_{s \subset\{1,\ldots,d\}}
    \int_{0_s}^{w_s} d \logit G_s(u_s) \nonumber \\
      & = \logit G(0) + \sum_{s \subset\{1,\ldots,d\}} \int_{0_s}^{\tau_s}
      \I(u_s \leq w_s) d \logit G_s(u_s).
\end{align}
The representation in equation~\ref{eq:hal1} may be approximated using a
discrete measure that puts mass on each observed $W_{s,i}$, denoted by
$\beta_{s,i}$. Letting $\phi_{s,i}(c_s)= \I(w_{s,i} \leq c_s)$, where $w_{s,i}$
are support points of $\logit G_s$, we have
\begin{equation*}
 \logit G_\beta = \beta_0 + \sum_{s \subset\{1,\ldots,d\}} \sum_{i=1}^{n}
 \beta_{s,i} \phi_{s,i},
\end{equation*}
where $\lvert \beta_0 \rvert + \sum_{s \subset\{1,\ldots,d\}} \sum_{i=1}^{n}
\lvert \beta_{s,i} \rvert$ is an approximation of the sectional variation norm
of $\logit G$. The loss-based highly adaptive lasso estimator $\beta_n$ may then
be defined as
\begin{equation*}
  \beta_{n, \lambda}= \argmin_{\beta: \lvert \beta_0 \rvert + \sum_{s
    \subset\{1,\ldots,d\}} \sum_{i=1}^{n} \lvert \beta_{s,i} \rvert <\lambda}
    P_n L(\logit G_\beta),
\end{equation*}
where $L(\cdot)$ is an appropriate loss function and $P_n f =n^{-1}
\sum_{i = 1}^n f(O_i)$. Denote by $G_{n,\lambda} \equiv G_{\beta_{n, \lambda}}$
the highly adaptive lasso estimate of $G_0$. When the functional nuisance
parameter is a conditional probability (e.g., the propensity score for a binary
treatment), log-likelihood loss may be used. Different choices of the tuning
parameter $\lambda$ result in unique highly adaptive lasso estimators; our goal
is to select a highly adaptive lasso estimator that allows the construction of
an asymptotically linear inverse probability weighted estimator of $\Psi(P_0)$.
We let $\lambda_n$ denote this data adaptively selected tuning parameter.


\section{Methodology}\label{methods}


We estimate the full data parameter $\Psi^F(P_X)$ using an inverse probability
weighted estimator $ \Psi (P_n,  G_{n})$, which is a solution to the score
equation $P_n U_{G_{n}}( \Psi)=0$. That is,
\begin{align}
  {\Psi}(P_n, G_{n}) = n^{-1} \sum_{i=1}^n
  \frac{A_i Y_i}{G_{n}(A_i \mid W_i) }. \label{eq:ipw}
\end{align}
Alternatively, a stabilized inverse probability weighted estimator may be
defined as the solution to $n^{-1} \{A_i (Y_i - \Psi(P))\} / \{G_{n}(A_i \mid
W_i) \} = 0$. The consistency and convergence rate of these estimators relies on
the consistency and convergence rate of the estimator $G_n$. While
finite-dimensional (i.e., parametric) models are often utilized to construct the
propensity score estimator $G_n$, it has been widely conceded that such models
are not sufficiently flexible to provide a consistent estimator of the nuisance
parameter $G_0$. Consequently, corresponding confidence intervals for
$\Psi^F(P_X)$ will have coverage tending to zero asymptotically. Flexible, data
adaptive regression techniques may be used to improve the consistency of $G_n$
for $G_0$; however, establishing the asymptotic linearity of the resultant
inverse probability weighted estimator $\Psi(P_n, G_n)$ can prove challenging.
Specifically,
\begin{align}\label{eq:asympl}
  \Psi(P_n, G_{n}) -{\Psi}(P_0, G_{0}) = &P_n U_{G_{n}}(\Psi) -
    P_0 U_{G_{0}}( \Psi) \nonumber \\
  = &(P_n - P_0) U_{G_{0}}( \Psi) + P_0 \{U_{G_{n}}(\Psi) -
   U_{G_{0}}( \Psi)\} \nonumber \\
  & + (P_n - P_0) \{U_{G_{n}}(\Psi) - U_{G_{0}}(\Psi)\}.
\end{align}
Assuming $U_G( \Psi)$ is c\`{a}dl\`{a}g with a universal bound on the
sectional variation norm, it can be shown that $(P_n - P_0) \{U_{G_{n}}(\Psi)
- U_{G_{0}}(\Psi)\} = o_p(n^{-1/2})$ for each $G$, relying only on standard
empirical process theory and the assumption of consistency.
Consequently, the asymptotic linearity of our inverse probability weighted
estimator relies on the asymptotic linearity of $P_0 \{U_{ G_{n}}( \Psi)
- U_{G_{0}}(\Psi)\}$.
Since data adaptive regression techniques have a rate of convergence slower than
$n^{-1/2}$, the bias of $P_0 \{U_{ G_{n}}(\Psi) - U_{ G_{0}}(\Psi)\}$ will
dominate the right-hand side of equation~\ref{eq:asympl}.

To show that asymptotic linearity of ${\Psi}(P_n, G_{n})$ can be established
when $G$ is estimated using a properly tuned highly adaptive lasso, we introduce
Lemma~\ref{lem:halipw1}, which is an adaptation of Theorem 1 of
\citet{vdl2019efficient}.
\begin{lemma}\label{lem:halipw1}
Let $G_{n,\lambda_n}$ be a highly adaptive lasso estimator of $G$ using
$L_1$-norm bound $\lambda_n$. Choosing $\lambda_n$ such that
\begin{align}\label{eq:basis}
    \min_{(s,j) \in \mathcal{J}_n} {\bigg \Vert} P_n \frac{d}{d\logit
    G_{n,\lambda_n}}
    L(\logit G_{n,\lambda_n}) (\phi_{s,j}) {\bigg \Vert} = o_p(n^{-1/2}),
\end{align}
where $L(\cdot)$ is log-likelihood loss and $\mathcal{J}_n$ is a set of indices
for the basis functions such that $\beta_{n,s,j} \neq 0$. Let $D(f,G_{n})
= f \cdot (A - G_{n})$. Here, $f$ is c\`{a}dl\`{a}g with finite sectional
variation norm, and we let $\tilde{f}$ be a projection of $f$ onto the linear
span of the basis functions $\phi_{s,j}$ in $L^2(P)$, where $\phi_{s,j}$
satisfies condition (\ref{eq:basis}). Assuming $\lVert f - \tilde{f}
\rVert_{2,P_0} = O_p(n^{-1/4})$, it follows that $P_n D(\tilde{f},G_{n}) =
o_p(n^{-1/2})$ and $P_n D({f},G_{n}) = o_p(n^{-1/2})$
where $\lVert f - \tilde{f} \rVert_{2, P_0}^2 = \int (f - \tilde{f})^2(o)
dP_0(o)$.
\end{lemma}

In condition (\ref{eq:basis}), $d / d \logit G_{n,\lambda_n} \{L(\logit
G_{n,\lambda_n}) (\phi_{s,j})\}$ is $d / d\epsilon \{L(\logit G_{n,\lambda_n}
+ \epsilon \phi_{s,j})\}$, denoting the directional derivative of the loss along
the path $\logit G^\epsilon_{n,\lambda_n} = \logit G_{n,\lambda_n} + \epsilon
\phi_{s,j}$. Under log-likelihood loss,
\begin{align*}
 \frac{d}{d\epsilon} L(\logit G^\epsilon_{n,\lambda_n}) (\phi_{s,j})
 \bigg\rvert_{\epsilon=0} &= (A - G_{n,\lambda_n}) \phi_{s,j}.
\end{align*}
Condition (\ref{eq:basis}) implies that those features $\phi_{s,j}$ that make
only a small change in the loss function will, on average, be included, thus
undersmoothing the fit.
In Theorem~\ref{th:halipw1}, we show that the use of an undersmoothed highly
adaptive lasso in estimating the nuisance parameter $G$ results in inverse
probability weighted estimators that are asymptotically linear and efficient in
the nonparametric model. This requires two assumptions:
\begin{assumption}\label{ass:cadlag}
  Let $Q_0(1,W)= \E(Y^1 \mid W)$ and $G_0(W)$ be c\`{a}dl\`{a}g with finite
  sectional variation norm.
\end{assumption}
\begin{assumption}\label{ass:proj}
  Let $\tilde{f}$ be the projection of $f = Q_0(1,W) / G_0$ onto a linear
  span of basis functions $\phi_{s,j}$ in $L^2(P)$, for $\phi_{s,j}$
  satisfying condition (\ref{eq:basis}). Then, $\lVert f - \tilde{f}
  \rVert_{2,P_0} = O_p(n^{-1/4})$.
\end{assumption}

Since the set of c\`{a}dl\`{a}g functions with finite sectional variation norm
contains a rich variety of functional forms, assumption~\ref{ass:cadlag} is mild
in that it would be expected to hold in nearly any practical application. Let's
now consider assumption~\ref{ass:proj}. This assumption states that the degree
of undersmoothing needs to be such that the generated basis functions in the
highly adaptive lasso fit of $G_n$ are sufficient to approximate $f$ within an
$n^{-1/4}$ neighborhood of $f$ (i.e., $\lVert f - \tilde{f} \rVert_{2,P_0}
= O_p(n^{-1/4})$). We know that, even without undersmoothing, these basis
functions are sufficient to approximate $\logit G_0$ within an $n^{-1/3}$
neighborhood (even when the coefficients are estimated). Let $f = Q_0(1,W)
/ G_0$ and suppose that $df_s/d\logit G_s < \infty$ for all sections
$s\subset\{0,1,\ldots,d\}$.
It follows that
\begin{align*}
  f(w) &=  f(0) + \sum_{s \subset \{1, \ldots, d \}}
     \int_{0_s}^{\tau_s} \I(u_s \leq w_s) df_{s}(u_s) \\
   &= f(0) + \sum_{s \subset \{1, \ldots, d \}} \int_{0_s}^{\tau_s}
     \I(u_s \leq w_s) \frac{d f_{s}(u_s)}{d \logit G_{s}(u_s)}
     d \logit G_{s}(u_s).
\end{align*}
Under this assumption, the set of basis functions needed to approximate
$\logit G$ are also sufficient to approximate $f$.
This implies that when $\logit G_0$ has similar complexity to $f$,
assumption~\ref{ass:cadlag} may hold without undersmoothing, where function
complexity is measured by the support set for the knot points of the basis
functions. On the other hand, when $G_0$ is a simple function (e.g., in
randomized controlled trials), undersmoothing is more likely to be needed so
that the undersmoothed $d\logit G_n$ has a rich enough support to approximate
$f$. In general, as $f$ becomes more complex relative to $G_0$, more
undersmoothing would be required. We examine this phenomenon in
Section~\ref{supp:addl_sims} of the Supplementary Material. In our simulations
we observe that even in the extreme case that $G_0(W)=0.5$ and $Q_0$ is a real
function of $W$, undersmoothing still improves the efficiency of inverse
probability weighted estimator based on the highly adaptive lasso.


\begin{theorem}\label{th:halipw1}
Let $G_{n,\lambda_n}$ be a highly adaptive lasso estimator of $G_0$ using bound
on the $L_1$-norm equal to $\lambda_n$, where $\lambda_n$ is a data-dependent
parameter chosen to satisfy condition (\ref{eq:basis}).
Under assumptions~\ref{ass:cadlag} and~\ref{ass:proj}, the estimator
$\hat{\psi} = \Psi(P_n, G_{n,\lambda_n})$ will be asymptotically efficient with
influence function
\begin{equation*}
  \hat{\psi} - \psi_0 = P_n \{U_{G_0}(\Psi) - D_{\text{CAR}}(P_0) \} +
    o_p(n^{-1/2}),
\end{equation*}
where $\psi_0=\Psi(P_0)$.
\end{theorem}

Intuitively, Theorem~\ref{th:halipw1} states that when the highly adaptive lasso
estimator $G_{n,\lambda_n}$ is properly undersmoothed, the resultant estimate
will include a rich enough set of basis functions to approximate any arbitrary
c\`{a}dl\`{a}g function with finite sectional variation norm (as per
Lemma~\ref{lem:halipw1}). With respect to the asymptotic linearity result,
condition (\ref{eq:basis}) implies that the chosen set of basis functions must
be sufficient to solve the efficient influence function equation, that is, $P_n
D_{\text{CAR}}(G_{n,\lambda_n}, Q_0) = o_p(n^{-1/2})$. A complete proof of this
result is given in Section~\ref{supp:proofs} of the Supplementary Material.

\section{Estimation}\label{estim}

\subsection{Cross-fitted inverse probability weighting
estimator}\label{estim_cv}

In order to circumvent the requirement that initial estimates of nuisance
parameters constructed by data adaptive regression fall in a Donsker class,
cross-fitting may be used to establish asymptotic linearity of the
resultant estimator~\citep{klaassen1987consistent}. Thus, cross-fitting may
allow for the relaxation of the assumption of a finite sectional variation norm
for both $Q_0$ and $G_0$ (i.e., assumption~\ref{ass:cadlag}). Even when $G_0$
falls in a Donsker class (e.g., when the selected $L_1$-norm remains bounded),
estimating the propensity score $G_n$ using $V$-fold cross-fitting can improve
the finite-sample performance of our estimators.


To employ $V$-fold cross-fitting, split the data, uniformly at random, into $V$
mutually exclusive and exhaustive sets of size approximately $n V^{-1}$. Denote
by $P_{n,v}^0$ the empirical distribution of a training sample and by
$P_{n,v}^1$ the empirical distribution of a validation sample. For a given
$\lambda$, exclude a single (validation) fold of data and fit the highly
adaptive lasso using data from the remaining $(V-1)$ folds; use this model to
estimate the propensity scores for observational units in the holdout
(validation) fold. Repeat this process $V$ times, such that holdout estimates of
the propensity score are available for all observational units. The cross-fitted
inverse probability weighted estimator $\widehat{\Psi}(P_{n,v}^1,
G_{n,\lambda})$ is the solution to $ V^{-1} \sum_{v=1}^V P_{n,v}^1
U_{G_{n,\lambda,v}}(\Psi)=0$,
where $G_{n,\lambda,v}(A \mid W)$ is the estimate of $G_{0}(A \mid W)$ applied
to the training sample for the v\textsuperscript{th} sample split for a given
$\lambda$.

Theorem~\ref{supp:th:halipwcross}, found in the Supplementary Material, shows
that the cross-fitted inverse probability weighted estimator is asymptotically
linear, allowing for the sectional variation norm of $G_n$ to diverge as $n$
increases, thus relaxing assumption~\ref{ass:cadlag}. In our numerical
experiments, presented in Section~\ref{sim} and Section~\ref{supp:addl_sims} of
the Supplementary Material, we find that a particular degree of undersmoothing
keeps the selected $L_1$-norm bounded as $n$ increases, across a diversity of
scenarios. Consequently, we view cross-fitting primarily as providing a finite
sample improvement.

\subsection{Undersmoothing in practice}\label{estim_usm}

Undersmoothing is crucial for both asymptotic linearity and efficiency of our
proposed estimators. Our theoretical results show that targeted undersmoothing
of the highly adaptive lasso estimator of $G$ can result in an inverse
probability weighted estimator $\hat{\psi}$ that is a solution to the efficient
influence function equation. In practice, an $L_1$-norm bound for an estimate of
$G$ may be obtained such that
\begin{equation}\label{eq:udcar}
  \lambda_n = \argmin_{\lambda} \bigg\lvert V^{-1} \sum_{v=1}^V P_{n,v}^1
  D_{\text{CAR}}(G_{n,\lambda,v}, Q_{n,v}) \bigg\rvert,
\end{equation}
where $Q_{n,v}$ is a cross-validated highly adaptive lasso estimate of
$Q_0(1,W)$ with the $L_1$-norm bound based on the global cross-validation
selector. For a general censored data problem and inverse probability of
censoring weighted highly adaptive lasso estimator, in certain complex settings,
the derivation of the efficient influence function can become involved. This
arises, for example, in longitudinal settings with many decision points. For
such settings, alternative criteria that do not require knowledge of the
efficient influence function may prove useful. To this end, we propose the
criterion:
\begin{equation}\label{eq:score}
 \lambda_n = \argmin_{\lambda}  V^{-1} \sum_{v=1}^V \left[ \sum_{(s,j) \in
 \mathcal{J}_n} \frac{1}{ \lVert \beta_{n,\lambda,v} \rVert_{L_1}}
 \bigg\lvert P_{n,v}^1 \tilde S_{s,j}(\phi,G_{n,\lambda,v})
 \bigg\rvert \right],
\end{equation}
in which $\lVert \beta_{n,\lambda} \rVert_{L_1} = \lvert \beta_{n,\lambda,0}
\rvert + \sum_{s \subset\{1, \ldots, d\}} \sum_{j=1}^{n} \lvert
\beta_{n,\lambda,s,j} \rvert$ is the $L_1$-norm of the coefficients
$\beta_{n,\lambda,s,j}$ in the highly adaptive lasso estimator $G_{n,\lambda}$
for a given $\lambda$, and $\tilde S_{s,j}(\phi, G_{n,\lambda,v}) =
\phi_{s,j}(W) \{A_{} - G_{n,\lambda,v}(1 \mid W)\}\{G_{n,\lambda,v}
(1 \mid W)\}^{-1}$. This score-based criterion leverages a general
characteristic of canonical gradients: propensity score terms always appear in
the denominator. Per Lemma~\ref{lem:halipw1}, enough basis functions must be
generated such that $P_n D(\tilde{f}, G_n) = o_p(n^{-1/2})$, for $\tilde{f}
= \sum_{(s,j) \in \mathcal{J}_n} \alpha_{s,j} \phi_{s,j}$ (i.e., linear
approximation of $\{G_0(W)\}^{-1} Q_0(1,W)$) and a particular vector $\alpha$.
While this could be achieved by solving all possible score equations $P_n
S_{s,j} (\phi,G_{n,\lambda,v})=o_P(n^{-1/2})$ (where $ S_{s,j}(\phi,
G_{n,\lambda,v}) = \phi_{s,j}(W) \{A_{} - G_{n,\lambda,v}(1 \mid W)\}$), such an
approach is not feasible in finite samples. Instead, our approach allows for
increasing $\lambda$ --- thereby successively solving as many score equations as
possible for a given sample --- until a desired tradeoff between decreasing the
score $S_{s,j}(\phi,G_{n,\lambda,v})$ (i.e., equation~\ref{eq:score}) and
increasing the variance of the weight function $\{G_{n,\lambda,v}(1 \mid
W)\}^{-1}$ is achieved. This corresponds to a bias-variance tradeoff for our
functional parameter. Another key component of our score criterion is the
$L_1$-norm $\lVert \beta_{n,\lambda} \rVert_{L_1}$. Under
assumption~\ref{ass:cadlag}, as $\lambda$ increases, the $L_1$-norm increases,
but its rate of increase diminishes as $\lambda$ diverges. Hence, at a certain
point in the grid of $\lambda$, decreases in
$S_{s,j}(\phi,G_{n,\lambda,v})/\lVert \beta_{n,\lambda} \rVert_{L_1}$ are
insufficient for satisfying equation~\ref{eq:score}, which starts increasing on
account of $\{G_{n,\lambda,v}(1 \mid W)\}^{-1}$.

In both of the proposed undersmoothing criteria, the series of propensity score
models based on the highly adaptive lasso is constructed as follows. First, an
initial model is fit via global cross-validation (to choose a starting value
$\lambda_{\textsc{cv}}$). Next, undersmoothed models are constructed by
weakening the restriction placed on the $L_1$-norm (i.e., $\lambda \geq
\lambda_{\textsc{cv}}$). Then, the value of $\lambda$ is increased until the
target criterion is satisfied, allowing a particular model in the sequence to be
selected.


\subsection{Stability under near-violations of positivity}\label{trunc}

In practice, the estimated propensity score may fall close to the boundaries of
the unit interval. In such cases, the assumption of positivity may be nearly
violated, resulting in large or unstable estimates of the inverse probability
weights required for estimator construction. In such situations, undersmoothing
may induce further instability by pushing propensity score estimates closer
still to the unit interval boundaries. That is, even achieving the degree of
undersmoothing required to ensure asymptotic linearity may result in inflating
the variance of the resultant inverse probability weighted estimator,
compromising its efficiency. To mitigate this tradeoff, we propose truncation of
propensity score estimates. For a given positive constant $\kappa$, truncation
sets all propensity score estimates lower than $\kappa$ or greater than
$1-\kappa$ to $\kappa$ and $1-\kappa$, respectively.

With only slight modification, the previously proposed undersmoothing criteria
may be used in selecting an optimal truncation level $\kappa$. To wit, the
selectors given in equations~\ref{eq:udcar} and~\ref{eq:score} may be
straightforwardly extended to their $\kappa$-truncated variants:
\begin{align}
  (\lambda_n, \kappa) &= \argmin_{\lambda, \kappa} \bigg\lvert V^{-1}
  \sum_{v=1}^V  P_{n,v}^1 D_{\text{CAR}}(G_{n, \lambda, v, \kappa},
     Q_{n,v}) \bigg\rvert,  \label{eq:udcar2} \\
    (\lambda_n, \kappa) & = \argmin_{\lambda,\kappa}  V^{-1}
      \sum_{v=1}^V \left[ \sum_{(s,j) \in \mathcal{J}_n}
      \frac{1}{\lVert \beta_{n,\lambda,v} \rVert_{L_1}} \bigg\lvert
      P_{n,v}^1 \tilde S_{s,j}(\phi,G_{n,\lambda,v,\kappa})
      \bigg\rvert \right],
\label{eq:score2}
\end{align}
where $G_{n, \lambda, v, \kappa}$ is the truncated propensity score estimate for
a given $\lambda$ and $\kappa$, and $\tilde S_{s,j}(\phi,
G_{n,\lambda,v,\kappa}) = \phi_{s,j}(W) \{A_{} - G_{n,\lambda,v,\kappa}(1 \mid
W)\}\{G_{n,\lambda,v,\kappa}(1 \mid W)\}^{-1}$.

\section{Numerical Studies}\label{sim}

The practical performance of our proposed inverse probability weighted
estimators was assessed in several simulation studies. We present two of these
studies in the sequel, with three additional scenarios discussed in
Section~\ref{supp:addl_sims} of the Supplementary Material. In the present two
scenarios, we assess the performance of our inverse probability weighted
estimators against alternatives based on correctly specified parametric models
for the propensity score, illustrating that estimators based on undersmoothing
of the highly adaptive lasso can be made unbiased and efficient.

In both of the following scenarios, $W_1 \sim \text{Uniform}(-2, 2)$, $W_2 \sim
\text{Normal}(\mu = 0, \sigma = 0.5)$, $\epsilon \sim \text{Normal}(\mu = 0,
\sigma = 0.1)$, and $\expit(x) = \{1 + \exp(-x)\}^{-1}$. In each setting, we
sample $n \in \{1000, 2000, 3000, 5000 \}$ independent and identically
distributed observations, applying each estimator to the resultant data. This
was repeated $200$ times. In both scenarios, the true propensity score $G_0$ is
bounded away from zero (i.e., $0.15 < G_0$); thus, the positivity assumption
holds. In both scenarios, the true treatment effect is zero.

In the first scenario, $A \mid W \sim \text{Bernoulli}\{\expit(0.75 W_1 + 0.5
W_2)\}$ and $Y \mid A, W = 0.5 W_1 - 2/3 W_2 + \epsilon$. As both models are
linear, parametric inverse probability weighted estimators are expected to be
unbiased. In the second scenario, $A \mid W \sim \text{Bernoulli}\{\expit(0.5
W_2^2 - 0.5\exp(W_1 / 2))\}$ and $Y \mid A, W = 2 W_1 -2 W_2^2 + W_2 + W_1 W_2
+ 0.5 + \epsilon$. Due to nonlinearity of the propensity score model, the
parametric inverse probability weighted estimator would be expected to exhibit
bias while our undersmoothed inverse probability weighted estimators ought to be
unbiased and efficient.

We consider different undersmoothing criteria including the minimizer of
$D_{\text{CAR}}$ (equation~\ref{eq:udcar}) and the alternative score-based
method (equation~\ref{eq:score}). Throughout, we use the highly adaptive lasso
\texttt{R} package, \texttt{hal9001}~\citep{coyle2019hal9001}, considering basis
functions for up to all 2-way interactions of covariates in estimating propensity
scores and the outcome $\E(Y \mid A=1, W)$. For comparison, we construct
propensity score estimates, using the highly adaptive lasso with
$\lambda$-selector based on cross-validation and a (parametric) logistic
regression model with main effect terms for $W_1$ and $W_2$. All models were fit
using 15-fold cross-validation. All numerical experiments were performed using
the \texttt{R} language and environment for statistical computing~\citep{R}.

\begin{figure}[H]
  \centering
  \includegraphics[scale=0.34]{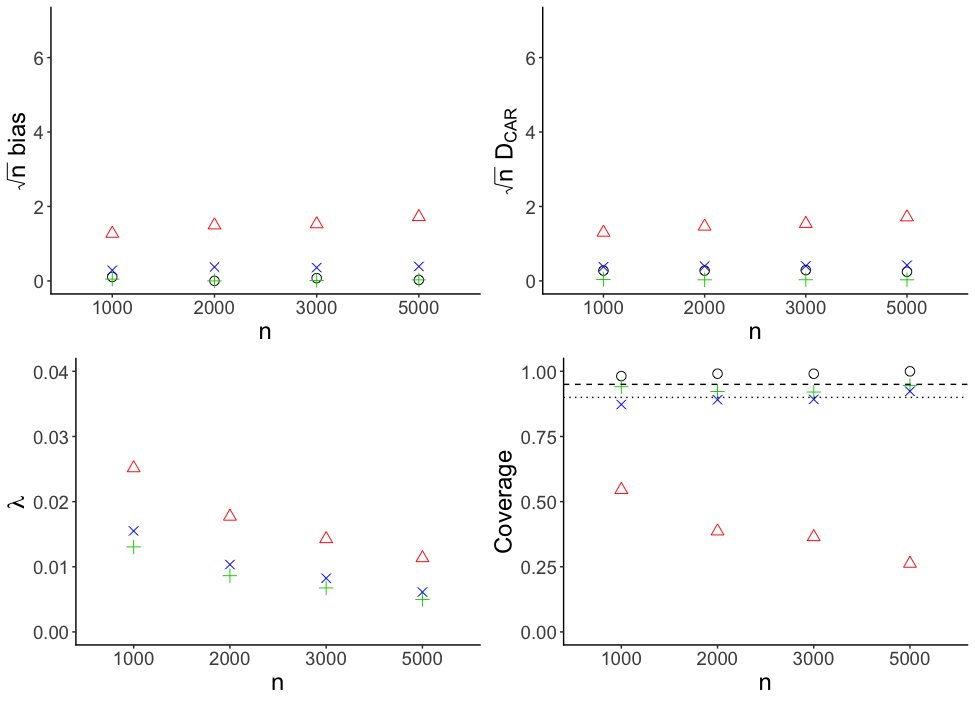}
  \caption{Comparative performance of inverse probability weighting variants in
  scenario 1. Circle: parametric; Triangle: nonparametric with cross-validated
  $\lambda$ selector; ``+'': $D_{\text{CAR}}$-based $\lambda$ selector; ``x'':
  score-based $\lambda$ selector.}
  \label{fig:lin}
\end{figure}

\begin{figure}[H]
  \centering
  \includegraphics[scale=0.33]{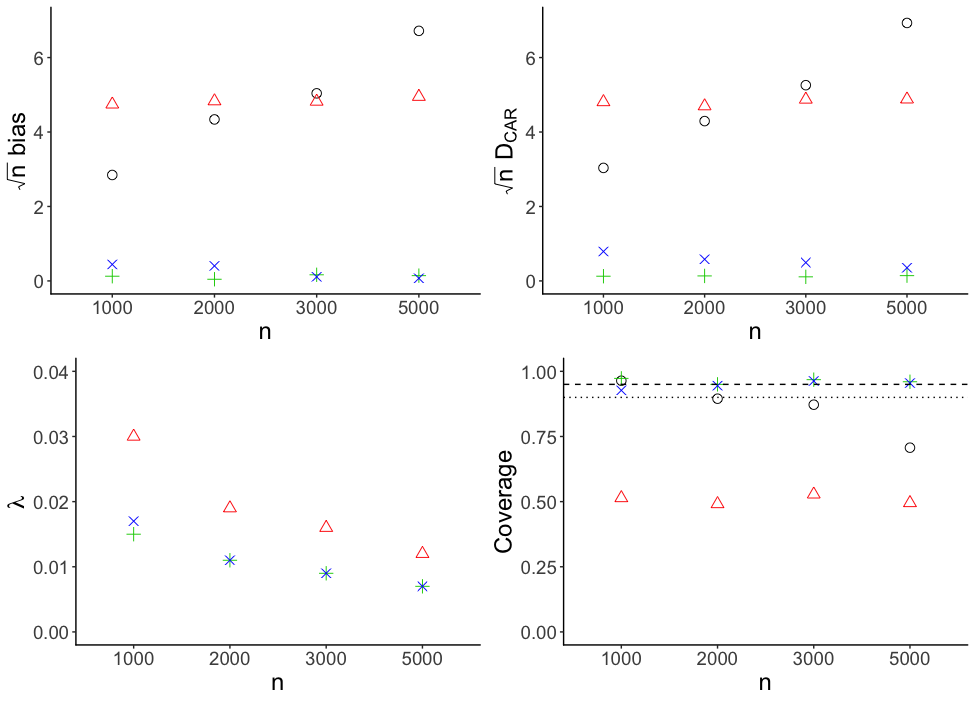}
  \caption{Performance of inverse probability weighting estimators in scenario
  2. Circle: parametric; Triangle: nonparametric with cross-validated $\lambda$
  selector; ``+'': $D_{\text{CAR}}$-based $\lambda$ selector; ``x'':
  score-based $\lambda$ selector.}
\label{fig:nonlin}
\end{figure}

Figures~\ref{fig:lin} and~\ref{fig:nonlin} display the results for scenarios
1 and 2, respectively.
The inverse probability weighted estimators based on undersmoothing of the
highly adaptive lasso outperform those based on the cross-validated highly
adaptive lasso in terms of both bias and efficiency, producing similar results
as the inverse probability weighted estimators based on the correctly specified
parametric model for the propensity score.

The first row of each figure presents the bias and the cross-validated mean of
$ D_{\text{CAR}}$ (both scaled by $n^{1/2}$) of the corresponding estimators,
where the latter is the objective function in equation \ref{eq:udcar}, and
expected to be nearly zero for estimators that solve the efficient influence
function equation. While the scaled bias and the cross-validated mean of
$D_{\text{CAR}}$ of the cross-validation-based selector diverges (triangle), the
undersmoothed highly adaptive lasso and the correctly specified parametric
models perform similarly. In terms of coverage, $D_{\text{CAR}}$-based criterion
achieves the nominal coverage rate of 95\%, even for smaller samples sizes,
while the cross-validation-based estimator (triangle) yields a poor coverage
rate of $\approx$50\%. The score-based undersmoothing selectors perform
reasonably well, producing inverse probability weighted estimators with coverage
rates $\approx$90\% for $n=1000$ and $\approx$95\% at larger sample sizes ($n
\geq 5000$). In scenario 2, where the parametric model of the propensity score
is misspecified, the parametric inverse probability weighted estimator performs
poorly, resulting in inverse probability weighted estimators with coverage rates
tending to zero asymptotically. In the same scenario, the score based selector
performs as well as the $D_{\text{CAR}}$-based selector producing estimators
with coverage rates $\approx$95\% for all the sample sizes considered. For both
scenarios, we additionally report the selected tuning parameter $\lambda$ based
on both the global cross-validation and undersmoothing selectors. Our results
illustrate that, as sample size increases, the selected value of the tuning
parameter stabilizes. Importantly, this observation implies that the
undersmoothing procedure does not lead to violations of the Donsker class
assumption.

We provide additional simulation studies in Section~\ref{supp:addl_sims} of the
Supplementary Material, in which we examine the relative performance of our
proposed estimators under differing outcome and propensity score models,
including settings corresponding to treatment randomization and observational
studies with positivity violations.

\section{Empirical Illustration}

\subsection{Overview and problem setup}

We now apply our proposed estimation strategy to assessing the effect of smoking
cessation on weight gain, using a subset of data from the National Health and
Nutrition Examination Survey Data I Epidemiologic Follow-up Study (NHEFS). As
per~\citet{hernan2020causal}, the NHEFS was jointly initiated by the National
Center for Health Statistics and the National Institute on Aging, in
collaboration with several other agencies of the United States Public Health
Service. The study was designed to investigate the impact of a variety of
clinical, nutritional, and behavioral factors on health outcomes including
morbidity and mortality. The subset of the NHEFS data we consider totaled
$n=1566$ cigarette smokers, all between the ages of 25 and 74; the data is
available at \url{https://hsph.harvard.edu/miguel-hernan/causal-inference-book}.
Each individual must have been present for a baseline visit and a follow-up
visit roughly 10 years later. Individual weight gain was measured as
a difference between baseline body weight and body weight at a follow-up visit;
moreover, individuals were classified as having been in the treatment group if
they reported having quit smoking prior to the follow-up visit and in the
control group otherwise. \citet{hernan2020causal} caution that this subset of
the NHEFS data could suffer from selection bias. As correcting for such a bias
is tangent to the illustration of our analytic approach, we forego standard
corrections, warning of this as a caveat of our demonstration. In practice, we
advocate the use of our strategy in tandem with censoring or selection bias
corrections, such as imputation or re-weighting by inverse probability of
censoring~\citep[e.g.,][]{carpenter2006comparison,seaman2012combining}.

\subsection{Estimation strategy}

We consider estimating the average treatment effect of smoking cessation on
weight gain in this subset of the NHEFS cohort ($n=1566$). A fairly rich set of
baseline covariates --- including sex, race, age, highest degree of formal
education, intensity of smoking, years of smoking, exercise habits, indicators
of an active lifestyle, and weight at study onset --- were considered as
potential baseline confounders of the relationship between smoking cessation and
weight gain. Constructing inverse probability weighted estimators for the
average treatment effect requires estimation of the propensity score, to model
the conditional probability of smoking cessation given potential baseline
confounders. An inverse probability weighted estimator for the average treatment
effect of smoking cessation may be constructed based on distinct estimators of
the respective treatment-specific means. We compare estimates of the average
treatment effect based on both parametric and nonparametric strategies for
estimating the propensity score, including
\begin{enumerate}[label=(\roman*),leftmargin=1cm]
  \item logistic regression with main terms for all baseline covariates;
  \item logistic regression with main terms for all baseline covariates and with
    quadratic terms for age, smoking intensity, years of smoking, and baseline
    weight; and
  \item the highly adaptive lasso with basis functions for all terms up to and
    including 3-way interactions between the baseline covariates, fit with
    $5$-fold cross-validation.
\end{enumerate}
The series highly adaptive lasso of propensity score models was constructed
weakening the restriction placed on the $L_1$-norm following an fit via global
cross-validation.

\subsection{Results}

We apply each of the inverse probability weighted estimators to recover the
average treatment effect of smoking cessation on weight gain, controlling for
possible confounding by the baseline covariates previously enumerated.
Table~\ref{tab:nhefs_ate} summarizes the results. Generally, estimates of the
average treatment effect were similar across the two classes of inverse
probability weighted estimators. When the propensity score was estimated via
a main terms logistic regression model, the estimate was 3.32 (CI: [2.15,
4.49]); likewise, when a logistic regression model with several quadratic terms
was used, the estimate was 3.42 (CI: [2.24, 4.61]). By contrast, our
cross-fitted (10 fold) nonparametric inverse probability weighted estimators
based on the highly adaptive lasso produced estimates of 3.23 (CI: [2.21,
4.26]) and 3.38 (CI: [2.29, 4.48]), for the $D_{\text{CAR}}$ and score-based
variants, respectively.
Confidence intervals corresponding to the proposed nonparametric estimators are
about 12\% shorter than those obtained by the parametric estimators, due to the
relatively enhanced efficiency of our estimators. Since the form of the
canonical gradient is readily known for the average treatment effect, in this
case, the $D_{CAR}$-based estimator provides the most reliable estimate. We note
that the $D_{CAR}$-based estimate of the average treatment effect is lower in
magnitude than those recovered by parametric methods, suggesting that the impact
of smoking cessation on weight gain may perhaps be attenuated when the
propensity score is estimated with an approach that is much less likely to be
misspecified than the parametric models relied upon in standard practice.

\begin{table}[H]
  \caption{Estimates of the average treatment effect from inverse probability
  weighted estimators using propensity score models based on logistic
  regression and the highly adaptive lasso selected via the $ D_{\text{CAR}}$
  and score-based criteria.}
  \begin{table}[H]
\centering
\begin{tabular}{lrrr}
Estimator & Lower 95\% CI & Estimate & Upper 95\% CI \\
Highly adaptive lasso ($D_{\text{CAR}}$) & 2.21 & 3.23 & 4.26 \\
Highly adaptive lasso (Score) & 2.29 & 3.38 & 4.48 \\
Logistic regression (main terms only) & 2.15 & 3.32 & 4.49 \\
Logistic regression (with quadratic terms) & 2.24 & 3.42 & 4.61 \\
\end{tabular}
\end{table}

   \label{tab:nhefs_ate}
\end{table}



\section*{Acknowledgement}

We thank David Benkeser for helpful discussions and practical insights. This
work was partially supported by the National Institute on Drug Abuse, the
National Institute on Alcohol Abuse and Alcoholism, and the National Institute
of Allergy and Infectious Diseases (award number R01-AI074345) of the National
Institutes of Health.





\bibliographystyle{biometrika}
\bibliography{references}
\end{document}


\maketitle

\section{Theoretical results: Proofs of Lemma~\ref{main:lem:halipw1} and
Theorem~\ref{main:th:halipw1}}\label{proofs}


\begin{proof}[of Lemma~\ref{main:lem:halipw1}]
For simplicity of notation, let $G^\dagger = \logit G$. Define a set of score
functions generated by a path $\{1+\epsilon h(s,j)\} \beta_{n,s,j}$ for
a bounded vector $h$ as
\begin{equation*}
  S_h(G_{n,\lambda_n}) = \frac{d}{dG^\dagger_{n,\lambda_n}} L(
  G^\dagger_{n,\lambda_n}) \left\{ \sum_{(s,j)} h(s,j) \beta_{n,s,j} \phi_{s,j}
  \right\}.
\end{equation*}
When $L(\cdot)$ is the log-likelihood loss function,
$L(G) =  A \log \left(\frac{G}{1-G}\right) +\log(1-G)$
Thus, $L(G^\dagger) =  A \log G^\dagger +\log(G^\dagger -1)$ and
$S_h(G) =   (A - G_{n,\lambda_n}) \left\{\sum_{(s,j)} h(s,j)
\beta_{n,s,j} \phi_{s,j} \right\}$.
Let $r(h,G_{n,\lambda_n}) = \sum_{(s,j)} h(s,j) \lvert \beta_{n,s,j} \rvert$.
For small enough $\epsilon$,
\begin{align*}
 \sum_{(s,j)} \lvert \{1+\epsilon h(s,j)\} \beta_{n,s,j} \rvert &= \sum_{(s,j)}
   \{1 + \epsilon h(s,j)\} \lvert \beta_{n,s,j} \rvert \\
   &=\sum_{(s,j)} \lvert \beta_{n,s,j} \rvert + \epsilon r(h,G_{n,\lambda_n}).
\end{align*}
Hence,  for any $h$ satisfying $r(h,G_{n,\lambda_n})=0$, we have $P_n
S_h(G_{n,\lambda_n}) = 0$. 

Let $D\{f,G_{n,\lambda_n}\} = f \cdot (A - G_{n,\lambda_n})$, where $f$ is
a c\`{a}dl\`{a}g function with a finite sectional variation norm, and let
$\tilde{f}$ be an approximation of $f$ using the basis functions that satisfy
condition (\ref{main:eq:basis}). Then, $D\{\tilde{f}, G_{n,\lambda_n}\} \in
\{S_h(G_{n,\lambda_n}): \lVert h \rVert_{\infty} < \infty \}$. Thus, there
exists $h^{\star}$ such that $D(\tilde{f}, G_{n,\lambda_n}) = S_{h^{\star}}
(G_{n,\lambda_n})$; however, for this particular choice of $h^{\star}$,
$r(h,G_{n,\lambda_n})$ may not be zero. Now, define $h$ such that $ h(s,j)
= h^{\star}(s,j)$ for $(s,j) \neq (s^{\star}, j^{\star})$; $\tilde{h}
(s^{\star}, j^{\star})$ is defined such that
\begin{align}\label{eq:restr}
\sum_{(s,j) \neq (s^{\star},j^{\star})} h^{\star}(s,j) \lvert
 \beta_{n,s,j} \rvert +  h(s^{\star}, j^{\star})
 \lvert \beta_{n, s^{\star}, j^{\star}} \rvert = 0.
\end{align}
That is, $h$ matches $h^{\star}$ everywhere but for a single point
$(s^{\star}, j^{\star})$, where it is forced to take a value such that
$r(h,G_{n,\lambda_n})=0$. As a result, for such a choice of $h$, $P_n S_{h}
(G_{n,\lambda_n}) = 0$ by definition. Below, we show that $P_n
S_{h}(G_{n,\lambda_n}) - P_n D\{\tilde{f}(W), G_{n,\lambda_n}\} = o_p(n^{-1/2})$
which then implies that $P_n D\{\tilde{f}(W), G_{n,\lambda_n}\}
= o_p(n^{-1/2})$. We note that the choice of $(s^{\star}, j^{\star})$ is
inconsequential.
\begin{align*}
   P_n S_{h}(G_{n,\lambda_n}) - P_n D\{\tilde{f}, G_{n,\lambda_n}\} &= P_n
   S_{h}(G_{n,\lambda_n}) - P_n S_{h^*}(G_{n,\lambda_n}) \\ &=P_n \left\{
   \frac{d}{dG^\dagger_{n,\lambda_n}} L( G^\dagger_{n,\lambda_n})
   \left[\sum_{(s,j)} \left\{h(s,j) - h^{\star}(s,j)\right\} \beta_{n,s,j}
   \phi_{s,j} \right]  \right\} \\ &= P_n
   \left[\frac{d}{dG^\dagger_{n,\lambda_n}} L( G^\dagger_{n,\lambda_n}) \left\{
   h(s^{\star},j^{\star}) - h^{\star}(s^{\star},j^{\star})\right\}
   \beta_{n,s^{\star},j^{\star}} \phi_{s^{\star},j^{\star}} \right] \\ & = P_n
   \left[\frac{d}{dG^\dagger_{n,\lambda_n}} L( G^\dagger_{n,\lambda_n})
   \kappa(s^{\star},j^{\star}) \phi_{s^{\star},j^{\star}} \right]\\ &=
   \kappa(s^{\star},j^{\star}) P_n \left[\frac{d}{dG^\dagger_{n,\lambda_n}} L(
   G^\dagger_{n,\lambda_n}) \phi_{s^{\star},j^{\star}} \right],
\end{align*}
where the third equality follows from equation (\ref{eq:restr}) above with
\begin{equation*}
 \kappa(s^{\star},j^{\star}) = -\frac{\sum_{(s,j) \neq (s^{\star},j^{\star})}
 h^{\star}(s,j) \lvert \beta_{n,s,j} \rvert }{\lvert
 \beta_{n,s^{\star},j^{\star}} \rvert} \beta_{n,s^{\star},j^{\star}} -
 h^{\star}(s^{\star},j^{\star}) \beta_{n,s^{\star},j^{\star}}.
\end{equation*}
Moreover,
\begin{equation*}
 \left\lvert \kappa(s^{\star},j^{\star})  \right\rvert \leq \sum_{(s,j)}
 \lvert h^{\star}(s,j)  \beta_{n,s,j} \rvert.
\end{equation*}
Assuming $f$ has finite sectional variation norm, the $L_1$ norm of the
coefficients approximating $ f$ will be finite which implies that $\sum_{(s,j)}
 \lvert h^{\star}(s,j)  \beta_{n,s,j} \rvert$ is finite, and thus
$\lvert \kappa(s^{\star},j^{\star}) \rvert = O_p(1)$. Then,
\begin{align*}
 P_n S_{h}(G_{n,\lambda_n}) - P_n D(\tilde{f}, G_{n,\lambda_n}) &=
 O_p \left(P_n \left[\frac{d}{dG^\dagger_{n,\lambda_n}} L(
 G^\dagger_{n,\lambda_n})\phi_{s^{\star},j^{\star}} \right] \right) \\
 & = o_p(n^{-1/2}),
\end{align*}
where the last equality follows from the assumption that $\min_{(s,j) \in
\mathcal{J}_n } \lVert P_n \frac{d}{dG^\dagger_{n,\lambda_n}}
L(G^\dagger_{n,\lambda_n}) (\phi_{s,j}) \rVert = o_p(n^{-1/2})$ for $L(\cdot)$
being log-likelihood loss. As $P_n S_{{h}}(G_{n,\lambda_n}) = 0$, it
follows that $P_n D(\tilde{f},G_{n,\lambda_n}) = o_p(n^{-1/2})$. Using this
result, it may be shown that, $P_n D(f,G_{n,\lambda_n})= o_p(n^{-1/2})$ under
mild assumptions as well. We have
\begin{align}\label{eq:prooflem1}
  P_n D(\tilde{f}, G_{n,\lambda_n}) - P_n D(f,G_{n,\lambda_n}) &=
  P_n (\tilde{f} - f) \left\{ \frac{d}{dG^\dagger_{n,\lambda_n}}
  L(G^\dagger_{n,\lambda_n})\right\} \nonumber \\ &= P_n (\tilde{f} - f)
  \left\{\frac{d}{dG^\dagger_{0}} L( G^\dagger_{0})\right\} 
  + P_n (\tilde{f} - f) (G_{0} - G_{n,\lambda_n}) \nonumber\\ &= o_p(n^{-1/2})
  + (P_n - P_0) (\tilde{f} - f)(G_{0} - G_{n,\lambda_n})\nonumber \\ & \quad
  + P_0 (\tilde{f} - f) (G_{0} - G_{n,\lambda_n})\nonumber\\ &= o_p(n^{-1/2}).
\end{align}
In the third equality, we use standard empirical process theory, the fact that
$P_0 \{f - \tilde{f}\}^2 \rightarrow 0$, and a Donsker class assumption to show
that
\begin{align*}
  P_n (\tilde{f} - f)\left\{\frac{d}{dG^\dagger_{0}} L( G^\dagger_{0})\right\}
  &= (P_n-P_0) (\tilde{f} - f) \left\{\frac{d}{dG^\dagger_{0}}
  L(G^\dagger_{0})\right\} \\ &= o_p(n^{-1/2}).
\end{align*}
The last equality (\ref{eq:prooflem1}) follows from the convergence rate of the
highly adaptive lasso estimate (i.e., $\lVert G_{0} - G_{n,\lambda_n}
\rVert_{2,P_0} = o_p(n^{-1/4})$) and the rate assumption $\lVert f - \tilde{f}
\rVert_{2,P_0} = O_p(n^{-1/4})$.
\end{proof}

\begin{proof}[of Theorem~\ref{main:th:halipw1}]
By definition, we have
\begin{align*}
  \Psi(P_n, G_{n,\lambda_n}) - \Psi(P_0, G_{0}) = \{\Psi(P_n, G_{n,\lambda_n})
  - \Psi(P_n, G_{0})\} + \{\Psi(P_n, G_{0}) - \Psi(P_0, G_{0})\}.
\end{align*}
The term $\Psi(P_n, G_{0})$ corresponds to an inverse probability weighted
estimator with known $G$; thus, it is asymptotically linear with influence
function $U_{G_0}(\psi_0) = (AY / G_0) -\psi_0$ for $\psi_0 = \Psi(P_0)$.
The term in the first bracket requires further attention, as it involves an
estimate of the nuisance functional parameter. This term may be expressed as
\begin{align*}
  \Psi(P_n, G_{n,\lambda_n}) - \Psi(P_n, G_{0}) &= P_n
     \left\{U_{G_{n,\lambda_n}}( \Psi)-U_{G_{0}}( \Psi)\right\} \\
    & = (P_n - P_0) \left\{U_{G_{n,\lambda_n}}(\Psi)-U_{G_{0}}(\Psi)
    \right\} + P_0  \left\{U_{G_{n,\lambda_n}}( \Psi)-U_{G_{0}}(
    \Psi)\right\} \\
    & = o_p(n^{-1/2}) + P_0  \left\{U_{G_{n,\lambda_n}}(
    \Psi)-U_{G_{0}}( \Psi)\right\}
\end{align*}
The last equality follows from the assumption that $G_0$ has finite sectional
variation norm which implies that $\{U_{G_{n,\lambda_n}}(\Psi)-U_{G_{0}}
(\Psi)\}$ also has finite sectional variation norm. By using
standard empirical process theory and a Donsker class assumption, we have
$(P_n - P_0)\{U_{G_{n,\lambda_n}}(\Psi) - U_{G_{0}}(\Psi)\} = o_p(n^{-1/2})$.
Define $D_{\text{CAR}}(Q_0,G_0,G_{n,\lambda_n}) =  Q_0(1,w) \left(\frac{A -
G_{n,\lambda_n}} {G_{0}} \right)$; then, the last term on the right-hand side
of the equation above may be expressed
\begin{align*}
   P_0 \left\{U_{G_{n,\lambda_n}}(\Psi)-U_{G_{0}}( \Psi)\right\}
   &= P_0 \left\{G_{0} Q_0(1) \left(\frac{G_{0} - G_{n,\lambda_n}}
      {G_{n,\lambda_n}G_{0}} \right)\right\} \\
    &= P_0 \left\{ Q_0(1) \left(\frac{G_{0} - G_{n,\lambda_n}}{G_{0}}
      \right) \right\} + P_0 \left\{\frac{Q_0(1)}{G_{n,\lambda_n}}
      \left(G_{0} - G_{n,\lambda_n} \right)^2 \right\} \\
    &= P_0 \left\{ Q_0(1) \left(\frac{G_{0} - G_{n,\lambda_n} }{G_{0}}
      \right) \right\} + o_p(n^{-1/2})\\
     & = -(P_n - P_0) \left\{D_{\text{CAR}}(Q_0,G_0,G_{n,\lambda_n})\right\}
     - P_n \left\{D_{\text{CAR}}(Q_0,G_0,G_{n,\lambda_n}) \right\} \\
     & \hspace{2in}+ o_p(n^{-1/2}) \\
     & = -(P_n - P_0) \left\{D_{\text{CAR}}(P_0)
       \right\} - P_n \left\{D_{\text{CAR}}(Q_0,G_0,G_{n,\lambda_n}) \right\} \\
       & \hspace{2in}+ o_p(n^{-1/2}), \\
\end{align*}
where $D_{\text{CAR}}(P_0) = Q_0(1,w) \left(\frac{A - G_{0}}{G_{0}} \right)$.
The third equality follows from the convergence rate of the highly adaptive
lasso estimator (i.e., $\lVert G_{0} - G_{n,\lambda_n} \rVert_2 =
o_p(n^{-1/4})$)~\citep{vdl2017generally}. The last equality follows from the
Donsker class condition, yielding $(P_n - P_0) \left\{D_{\text{CAR}}(P_0) -
D_{\text{CAR}}(Q_0,G_0,G_{n,\lambda_n}) \right\} = o_p(n^{-1/2})$. Applying
Lemma~\ref{main:lem:halipw1} with $f = Q_0(1,w) / G_0$, it readily follows that
$P_n \left\{D_{\text{CAR}}(Q_0,G_0,G_{n,\lambda_n}) \right\} = o_p(n^{-1/2})$.



Finally, gathering all of the above terms, we have
\begin{align*}
  \Psi(P_n, G_{n,\lambda_n}) - \Psi(P_0, G_{0}) = (P_n - P_0)
  \left\{U_{G_0}(\Psi) - D_{\text{CAR}}(P_0) \right\} + o_p(n^{-1/2}),
\end{align*}
where $U_{G_0}(\Psi) = (AY / G_0) - \Psi$. Further, the canonical gradient of
$\Psi$ at $P$ is given by
\begin{equation*}
  D^{\star}(P) = U_G(\Psi) - \prod \{U_G(\Psi) \mid T_{\text{CAR}}\}.
\end{equation*}
Following~\citet{vdl2003unified}, $\prod \{U_G(D^F) \mid T_{\text{CAR}}\}$ is
given by $\E \{U_G(O,\Psi) \mid A = a, W\} - \E \{U_G(O,\Psi) \mid W \}$, which
may be expressed as
\begin{align*}
  D_{\text{CAR}}(P) =  \frac{A - G(A \mid W)}{G(A \mid {W})} Q(1, W).
\end{align*}
This completes the proof, establishing the asymptotic linearity and efficiency
of our proposed estimators.
\end{proof}

\begin{theorem}\label{th:halipwcross}
Let $G_{n,\lambda_n}$ be a cross-fitted highly adaptive lasso estimator of $G_0$
using bound on the $L_1$-norm equal to $\lambda_n$, where $\lambda_n$ is
a data-dependent parameter chosen to satisfy condition (\ref{main:eq:basis}).
Under the positivity assumption and Assumption ~\ref{main:ass:proj}, the
estimator $\hat{\psi}^{cf} = \Psi(P_{n,v}^1, G_{n,\lambda_n})$ will be
asymptotically efficient with influence function
\begin{equation*}
  \hat{\psi}^{cf} - \psi_0 = P_n \{U_{G_0}(\Psi) - D_{\text{CAR}}(P_0) \} +
    o_p(n^{-1/2}),
\end{equation*}
where $\psi_0=\Psi(P_0)$.
\end{theorem}

\begin{proof}
Define the cross-fitted inverse probability weighted estimator
$\widehat{\Psi}(P_{n,v}^1, G_{n,\lambda})$ is defined as the solution to
$\frac{1}{V} \sum_{v=1}^V P_{n,v}^1 U_{G_{n,\lambda,v}}(\Psi)=0$, where
$G_{n, \lambda, v}(A \mid W)$ is the  estimate of $G_{0}(A \mid W)$ applied to
the training sample for the $v^{\text{th}}$ sample split for a given $\lambda$.
Moreover, by Lemma~\ref{main:lem:halipw1}, for any given $v$, $P_{n,v}^1
D_{CAR}(G_{n,\lambda,v},Q_0) = o_p(n^{-1/2})$. Thus, for any finite $V$,
$\frac{1}{V} \sum_{v=1}^V P_{n,v}^1 D_{CAR}(G_{n,\lambda,v},Q_0) =
o_p(n^{-1/2})$. Let $D^{\star}(G, Q, \psi) = U_{G}(\psi) - D_{CAR}(G,Q)$
be the efficient influence function. Then,
\begin{align*}
  \hat \psi-\psi_0 &= -\frac{1}{V} \sum_{v=1}^V P_0
        D^{\star}(G_{n,\lambda,v},Q_0,\hat{\psi}) \\
  &=  \frac{1}{V} \sum_{v=1}^V (P_{n,v}^1-P_0)
        D^{\star}(G_{n,\lambda,v},Q_0,\hat \psi) -
   \frac{1}{V} \sum_{v=1}^V P_{n,v}^1 D^{\star}(G_{n,\lambda,v},
        Q_0, \hat{\psi}) \\
   &=  \frac{1}{V} \sum_{v=1}^V (P_{n,v}^1-P_0) D^{\star}(G_{n,\lambda,v},
        Q_0, \hat{\psi}) + o_p(n^{-1/2}).
\end{align*}
Conditioning on the training sample, the term $\frac{1}{V} \sum_{v=1}^V
(P_{n,v}^1 - P_0) D^{\star}(G_{n,\lambda,v},Q_0, \hat{\psi})$ is an empirical
mean of mean-zero random variables. Thus, under consistency of the estimator
$G_{n,\lambda}$ and the positivity assumption, for any given fold $v$, it
follows that $(P_{n,v}^1-P_0) D^{\star}(G_{n,\lambda,v},Q_0, \hat{\psi})
\rightsquigarrow N(0, \mathbb{V}\{D^{\star}(G_{0}, Q_0, \psi_0)\})$.
\end{proof}

\section{Additional Simulation Studies}\label{addl_sims}

To further evaluate the performance of our proposed procedures, we
interrogate our estimators in three additional sets of simulation studies. In
each scenario, we compare five variants of our inverse probability weighted
estimators to an alternative unadjusted estimator that considers only $\E Y$ in
the treatment group (i.e., $A = 1$). The inverse probability weighted estimators
are constructed using varied selectors for the highly adaptive lasso estimator
of the propensity score, including
\begin{enumerate}[label=(\roman*),leftmargin=1cm]
  \item global cross-validation to choose the $L_1$-norm;
  \item undersmoothing with minimization of the cross-validated mean of
    $D_{\text{CAR}}$ (as per equation~\ref{main:eq:udcar}), with and without
    $\kappa$-truncation; and
  \item solving the cross-validated score-based criterion arising (from
    equation~\ref{main:eq:score}), with and withou $\kappa$-truncation.
\end{enumerate}
Each of the inverse probability weighted estimators was constructed using 5-fold
cross-fitting, as discussed in Section~\ref{main:estim_cv}. In the sequel, we
consider scenarios across which the true treatment assignment mechanism ranges
from randomization to a complex function of the covariates $W$.

In the first scenario, the treatment assignment mechanism is randomized, such
that the baseline covariates have no impact on the probability of receiving
treatment. In the second scenario, we consider a treatment assignment mechanism
that includes interaction terms between baseline covariates and in which the
positivity assumption strongly holds. In the third and final scenario, we
consider a treatment mechanism that includes interaction terms between baseline
covariates and in which the probability of receiving treatment is extremely low
for certain strata of the covariates $W$.

In all sets of simulation studies, we consider estimation of $\Psi(P_0)
= \E_{P_0}\{Y(1)\}$, the counterfactual mean for having received treatment. In
each scenario, $n$ independent and identically distributed observations were
drawn from the given data-generating mechanism, for each of the sample sizes $n
\in \{100, 400, 900, 1600\}$. Throughout, $\epsilon \sim \text{Normal}(\mu = 0,
\sigma = 0.1)$ and $\expit(x) = \{1 + \exp(-x) \}^{-1}$. Each of the estimators
is compared in terms of their raw (unscaled) bias, scaled (by $n^{1/2}$) bias,
scaled (by $n$) mean squared error, and the coverage of 95\% Wald-style
confidence intervals. For each sample size, the experiment was repeated between
200 and 500 times, with estimator performance evaluated by aggregation across
these repetitions. All numerical experiments were performed using the \texttt{R}
language and environment for statistical computing~\citep{R}.

\subsection{Simulation \#1: Randomized controlled trial}\label{sim:1b}

\begin{figure}[H]
  \includegraphics[trim=4cm 0 0 0,scale=0.28]{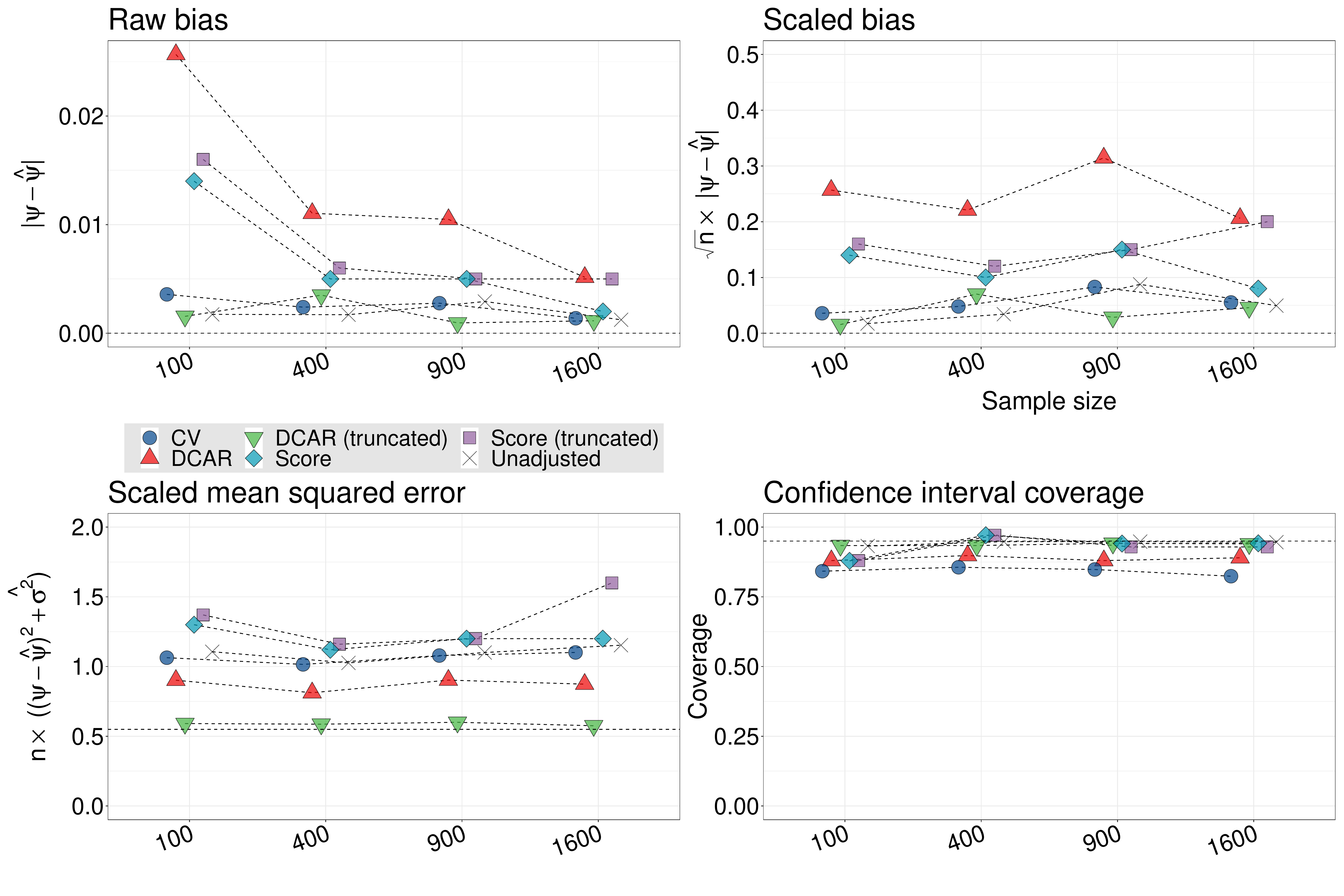}
  \caption{Relative performance of a simple unadjusted estimator and
  cross-fitted inverse probability weighted estimators based on the
  undersmoothing selectors and the cross-validation selector, in a randomized
  controlled trial.}
  \label{fig:dgp1b}
\end{figure}

We evaluate the inverse probability weighted estimators based on data from the
following data-generating mechanism:
$W_1 \sim \text{Uniform}(0.2, 0.8)$,
$W_2 \sim \text{Bernoulli}(p = 0.3)$;
$A \mid W \sim \text{Bernoulli}(p = 0.5)$; and
$Y \mid A, W = A \cdot (W_1 + W_2 + W_1 \cdot W_2) + (1 - A) \cdot W_1 +
\epsilon$.
The true value of the target parameter was numerically approximated, $\psi_0
= 0.95$. In this setting, it is expected that the unadjusted estimator be
unbiased, with mean squared error low enough that confidence intervals attain
the nominal coverage rate. By contrast, the undersmoothing-based estimators
would be expected to show low bias, with (scaled) mean squared error tending to
the efficiency bound. A comparison of the relative performance of the inverse
probability weighted estimators is presented in Figure~\ref{fig:dgp1b}.

Of the six estimators, all show negligible bias regardless of sample size. The
undersmoothing-based estimators that select the highly adaptive lasso propensity
score fit to minimize the cross-validated mean of $D_{\text{CAR}}$ provide the
best performance. In particular, the truncated variant of this estimator (green
triangle) uniformly achieves the lowest bias, has scaled mean squared error
nearest the efficiency bound, and results in confidence
intervals that cover at the nominal rate across all sample sizes. The
non-truncated variant of this estimator (red triangle) has the next lowest
scaled mean squared error but appears to suffer from a finite-sample bias that
compromises the coverage of its resultant confidence intervals.

The inverse probability weighted estimator that relies upon the cross-validation
selector (dark purple circle) has negligible bias, but fails to achieve the
efficiency bound or produce confidence intervals covering at the nominal rate;
thus, we conjecture its loss of efficiency to be due to a comparatively
increased variance. The estimator using the score-based selector without
truncation (cyan diamond) displays higher bias and lower efficiency than the
highly adaptive lasso-based estimator using the truncated $D_{\text{CAR}}$
selector, but this is not surprising, as our simulation experiments reported in
Section~\ref{main:sim} showed similarly compromised performance in relatively
small samples. Interestingly, the variant with truncation (purple square)
displays compromised performance (unlike in the case of the $D_{\text{CAR}}$
selector), causing an asymptotic bias that sharply degrades its efficiency. As
expected, the unadjusted estimator (``x'') shows negligible bias but makes
a different bias-variance tradeoff than the undersmoothing-based estimators. In
particular, the unadjusted estimator displays similar scaled mean squared error
to the estimator based on the cross-validation selector. Altogether, our
findings suggest that choosing the highly adaptive lasso estimator that
minimizes the $D_{\text{CAR}}$ term of the efficient influence function equation
while (allowing a degree of truncation) yields inverse probability weighted
estimators that are unbiased and more efficient than their standard counterparts
in these relatively limited sample sizes.


\subsection{Simulation \#2: Observational study without positivity
violations}\label{sim:1a}

\begin{figure}[H]
  \includegraphics[trim=4cm 0 0 0,scale=0.28]{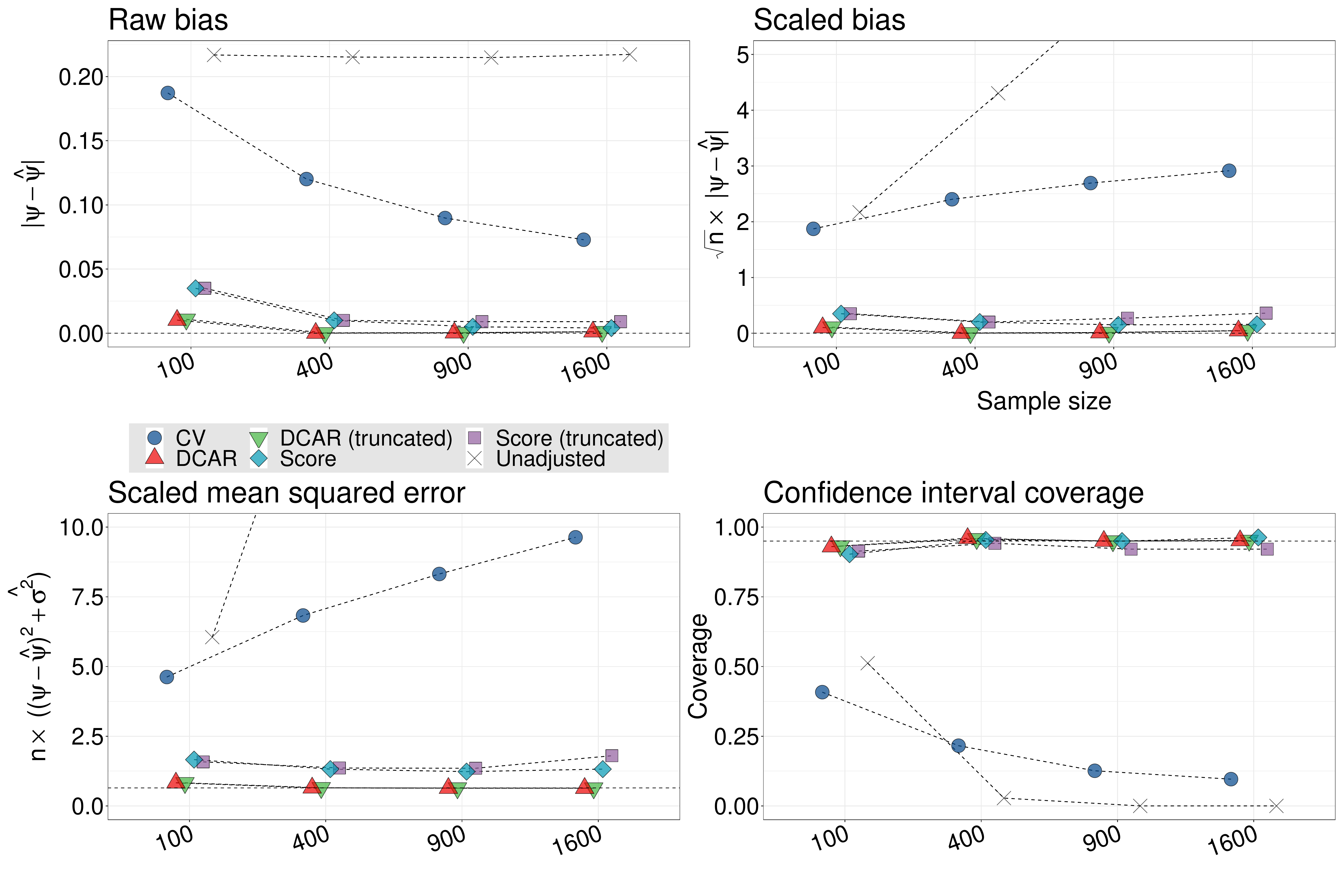}
  \caption{Relative performance of a simple unadjusted estimator and
  cross-fitted inverse probability weighted estimators based on the
  undersmoothing selectors and the cross-validation selector, in an
  observational study without positivity violations in the treatment mechanism.}
  \label{fig:dgp1a}
\end{figure}

In this setting, we evaluate the estimators based on data from the following
data-generating mechanism:
$W_1 \sim \text{Uniform}(0.2, 0.8)$,
$W_2 \sim \text{Bernoulli}(p = 0.6)$;
$A \mid W \sim \text{Bernoulli}(p = \expit(2W_1 - W_2 - W_1 \cdot W_2))$; and
$Y \mid A, W = A \cdot (W_1 + W_2 + W_1 \cdot W_2) + (1 - A) \cdot W_1 +
\epsilon$.
By numerical approximation, the true value of the target parameter was found to
be $\psi_0 = 1.40$. The treatment mechanism of this data-generating process is
not subject to positivity violations, with the probability of receiving
treatment conditional on covariates being bounded in $[0.31, 0.83]$. In this
setting, it would be expected that the unadjusted estimator would be quite
biased, with mean squared error growing with sample size and confidence interval
coverage approaching zero. By contrast, the undersmoothing-based estimators
would be expected to show low bias, with mean squared error asymptotically
tending to the efficiency bound. The estimator based on the global
cross-validation selector would also be expected to perform poorly in terms of
both bias and efficiency. The performance of the inverse probability weighted
estimators is compared in Figure~\ref{fig:dgp1a}.

Examination of Figure~\ref{fig:dgp1a} reveals that all four of the
undersmoothing-based estimators display desirable performance. As expected, the
estimators based on minimization of $D_{\text{CAR}}$ (red/green triangles)
provide the best finite-sample performance. As truncation makes no apparent
difference between these two estimator variants, we conjecture that the
truncation selector opts to avoid truncation, which is desirable in a setting
without positivity violations. The inverse probability weighted estimator
constructed from the score-based selector without truncation (cyan diamond)
achieves comparable performance, with very low bias and scaled mean squared
error stabilizing slightly above the efficiency bound. Importantly, this
estimator was not necessarily expected to provide desirable finite-sample
performance but does indeed do so, illustrating that our undersmoothing-based
estimators would have provided a viable option even without knowledge of the
exact form of the efficient influence function equation. The truncated variant
of this estimator (purple square) exhibits performance similar to its
non-truncated counterpart in smaller samples ($n \in \{100, 400\}$) but appears
to make a suboptimal tradeoff in larger samples, with slightly higher bias and
confidence intervals failing to cover at the nominal rate.

Still, three of the undersmoothing-based estimator variants allow the
construction of confidence intervals that cover at (nearly) the nominal rate,
regardless of sample size. By comparison, the estimator based on global
cross-validation (dark purple circle) displays considerable bias and mean
squared error, both increasing with sample size. Interestingly, the scaled bias
of this estimator appears to stabilize with increasing sample size; this
consistency suggests that the resultant estimator is asymptotically linear,
though it fails to be efficient. Confidence intervals based on this estimator
display very poor coverage, which decreases sharply with sample size. As
expected, the unadjusted estimator (``x'') fares poorly --- that is, it displays
a diverging bias, with similarly increasing mean squared error, and confidence
interval coverage that decreases to nearly zero by $n = 400$. These results
speak favorably of our proposed undersmoothing-based estimators, showing that
they may succesfully be used to construct unbiased and efficient estimators in
observational studies in which the treatment mechanism is not a simple linear
function of baseline covariates.


\subsection{Simulation \#3: Observational study with near-positivity
violations}\label{sim:2a}

\begin{figure}[H]
  \includegraphics[trim=4cm 0 0 0,scale=0.28]{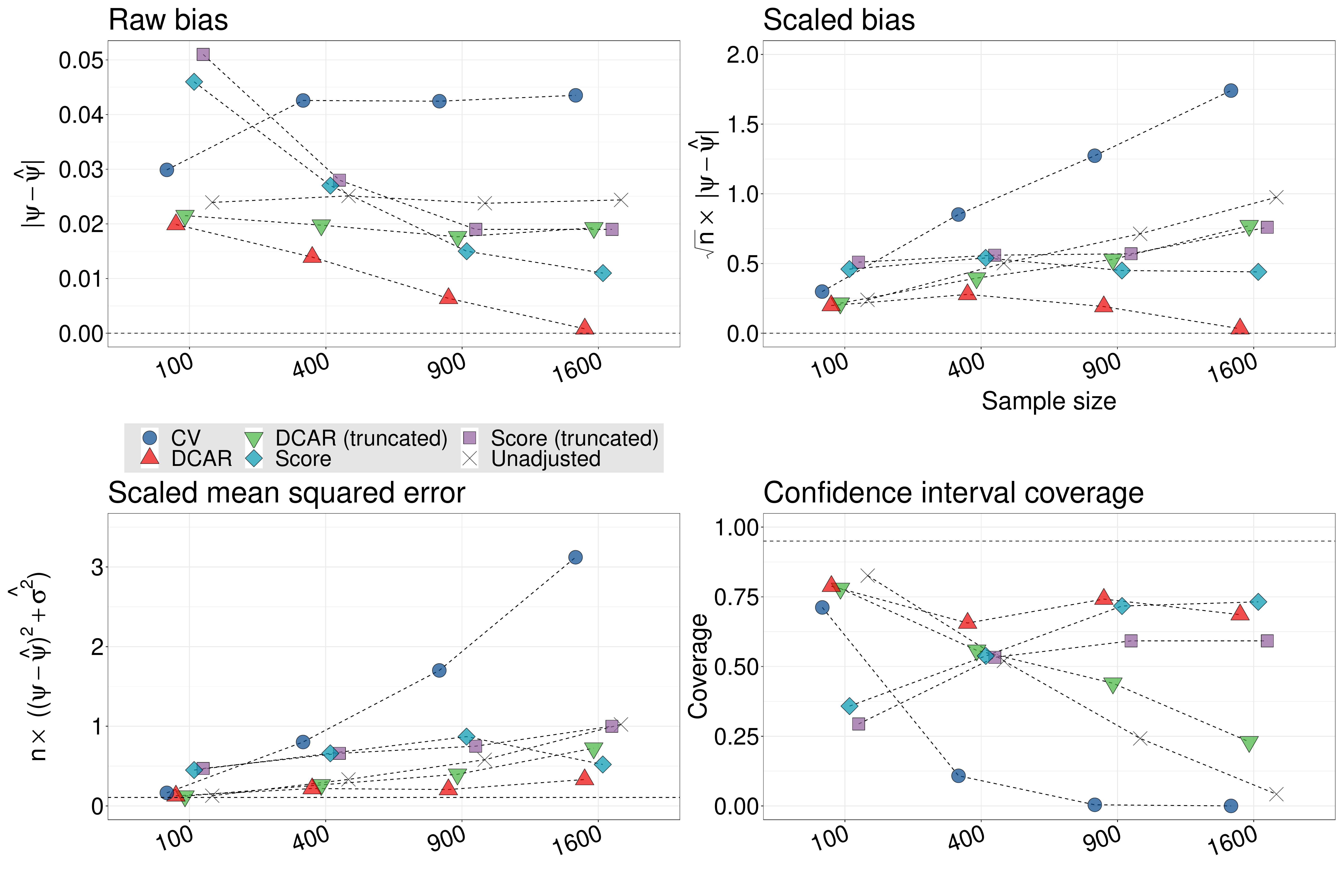}
  \caption{Performance of the unadjusted estimator and cross-fitted inverse
  probability weighted estimators based on the undersmoothing selectors and the
  cross-validation selector, in an observational study with positivity
  violations in the treatment mechanism.}
  \label{fig:dgp2a}
\end{figure}

In this setting, we evaluate the estimators based on data from the following
data-generating mechanism:
$W_1 \sim \text{Uniform}(0, 0.6)$,
$W_2 \sim \text{Bernoulli}(p = 0.05)$;
$A \mid W \sim \text{Bernoulli}(p = \expit(2W_1 - 4W_2 + W_1 \cdot W_2))$; and
$Y \mid A, W = A \cdot (W_1 + W_2) + (1 - A) \cdot W_1 + \epsilon$.
Numerical approximation was used to ascertain the true value of the target
parameter: $\psi_0 = 0.35$. The treatment mechanism of this data-generating
process is subject to positivity violations, with the probability of receiving
treatment conditional on covariates having a minimum of $0.018$. The unadjusted
estimator would be expected to be quite biased, with mean squared error growing
with sample size and confidence interval coverage converging to zero
asymptotically. By contrast, the undersmoothing-based estimators would be
expected to show low bias, with mean squared error asymptotically tending to the
efficiency bound. Such expectations might be particularly true of the truncated
$D_{\text{CAR}}$-based estimator variant, which may cope better with the
presence of destabilizing positivity violations. The estimator based on global
cross-validation would also be expected to perform poorly, failing to select an
$L_1$-norm appropriate for constructing an asymptotically linear estimator. The
performance of the inverse probability weighted estimators is presented in
Figure~\ref{fig:dgp2a}.

Most estimator variants perform quite poorly in this setting, though this is
expected of inverse probability weighted estimators. Since the treatment
mechanism includes severely destabilizing positivity violations, the inverse
probability of treatment weights can tend to very large values for a subset of
observed units, leading to bias and inefficiency in all estimators that
consistently estimate the propensity score. As expected, the unadjusted
estimator (``x'') and the estimator based on global cross-validation (dark
purple circle) display the worst performance, with asymptotically increasing
bias and mean squared error as well as confidence interval coverage tending to
zero. By comparison, the undersmoothing-based estimators perform better, though
not as well as might be desirable. In particular, the undersmoothing-based
estimator variant minimizes the cross-validated mean of $D_{\text{CAR}}$ and
utilizes truncation (green triangle) displays an asymptotically increasing bias,
suggesting that an optimal choice of the truncation parameter $\kappa$ was not
available to it. What's more, this bias contributes directly to increasing its
scaled mean squared error and adversely affects the coverage of its resultant
confidence intervals. As the truncation inherent in this estimator ought to
protect against the destabilizing bias caused by overly large inverse
probability of treatment weights, we surmise that pairing of the undersmoothing
and truncation selectors results in a suboptimal bias-variance tradeoff in this
case. Still, this estimator variant proves to be the second most efficient in
smaller sample sizes ($n < 1600$).

Somewhat contrary to expectations, the undersmoothing-based estimator without
truncation (red triangle) exhibits the best performance, with uniformly
negligible bias and mean squared error tending to the efficiency bound by $n
= 1600$. Unfortunately, the coverage of Wald-style confidence intervals based on
this procedure fail to attain the nominal rate of 95\%. Taken together with the
vanishing bias of this estimator, this relatively poor coverage suggests that
the estimator may suffer from instability in variance estimates in a handful of
instances. Surprisingly, the estimator constructed from the score-based selector
without truncation (cyan diamond) displays a bias that decreases with sample
size and appears promising asymptotically. This estimator variant displays
relatively good scaled mean squared error, tending to the efficiency bound at
$n = 1600$; confidence intervals based on this estimator provide increasing
coverage with sample size, comparable to those of the non-truncated
$D_{\text{CAR}}$-based estimator in larger samples ($n \in \{900, 1600\}$). In
particular, this increasing coverage suggests that confidence intervals of this
estimator variant may tend to coverage at the nominal rate asymptotically. The
$\kappa$-truncated variant of this estimator (purple square) generally displays
worse performance than its non-truncated counterpart, with uniformly higher bias
and scaled mean squared error. Interestingly, confidence intervals of this
estimator variant achieve better coverage than those of the truncated
$D_{\text{CAR}}$-based variant despite its lower efficiency, suggesting that
truncation may induce a higher variance in estimators derived from the
score-based selectors.

As expected, the unadjusted estimator (``x'') and the inverse probability
weighted estimator based on the cross-validation selector (dark purple circle)
provide the worst performance, with increasing scaled bias and mean squared
error, and constructed confidence intervals whose coverage degrades sharply with
increasing sample size. Altogether, this set of numerical experiments suggests
undersmoothing-based estimators may not be appropriate for settings in which the
true treatment mechanism is subject to positivity violations in the relatively
small sample sizes we consider. Still, both the non-truncated
$D_{\text{CAR}}$-based estimator and the score-based estimator exhibit
acceptable performance, though doubly robust estimators may prove a generally
better choice than inverse probability weighted estimators in settings with such
severe positivity violations. We leave further examination of these estimators,
and possible adjustments to the truncated estimator variants, as an area of
future investigation.

\bibliographystyle{biometrika}
\bibliography{references}